\documentstyle[sprocl]{article}

\input{psfig}

\begin{document}
\title{INTERPRETATIONS OF QUANTUM MECHANICS, AND
INTERPRETATIONS OF VIOLATION OF BELL'S INEQUALITY}

\author{WILLEM. M. DE MUYNCK}

\address{Theoretical Physics,
Eindhoven University of Technology,\\ POB 513, 5600 MB
Eindhoven, the Netherlands\\
E-mail: W.M.d.Muynck@tue.nl}


\maketitle\abstracts{ The discussion of the foundations of quantum
mechanics is complicated by the fact that a number of different
issues are closely entangled. Three of these issues are i) the
interpretation of probability, ii) the choice between realist and
empiricist interpretations of the mathematical formalism of
quantum mechanics, iii) the distinction between measurement and
preparation. It will be demonstrated that an interpretation of
violation of Bell's inequality by quantum mechanics as evidence of
non-locality of the quantum world is a consequence of a particular
choice between these alternatives. Also a distinction must be
drawn between two forms of realism, viz. a) realist
interpretations of quantum mechanics, b) the possibility of
hidden-variables (sub-quantum) theories.}

\section{Realist and empiricist interpretations of quantum mechanics}
\label{sec1}

In {\em realist} interpretations of the mathematical formalism of
quantum mechanics state vector and observable are thought to refer
to the microscopic object in the usual way presented in most
textbooks. Although, of course, preparing and measuring
instruments are often present, these are not taken into account in
the mathematical description (unless, as in the theory of
measurement, the subject is the interaction between object and
measuring instrument).

In an {\em empiricist} interpretation quantum mechanics is thought
to describe relations between input and output of a measurement
process. A state vector is just a label of a preparation
procedure; an observable is a label of a measuring instrument. In
an empiricist interpretation quantum mechanics is not thought to
describe the microscopic object. This, of course, does not imply
that this object would not exist; it only means that it is not
described by quantum mechanics. Explanation of relations between
input and output of a measurement process should be provided by
another theory, e.g. a hidden-variables (sub-quantum) theory. This
is analogous to the way the theory of rigid bodies describes the
empirical behavior of a billiard ball, or to the description by
thermodynamics of the thermodynamic properties of a volume of gas,
explanations being relegated to theories describing the
microscopic (atomic) properties of the systems.

Although a term like `observable' (rather than `physical
quantity') is evidence of the empiricist origin of quantum
mechanics (compare Heisenberg\cite{Heis25}), there has always
existed a strong tendency toward a realist interpretation in which
observables are considered as properties of the microscopic
object, more or less analogous to classical ones. Likewise, many
physicists use to think about electrons as wave packets flying
around in space, without bothering too much about the
``Unanschaulichkeit'' that for Schr\"odinger\cite{Schroed35} was
such a problematic feature of quantum theory. Without entering
into a detailed discussion of the relative merits of either of
these interpretations (e.g. de Muynck\cite{dM95}) it is noted here
that an empiricist interpretation is in agreement with the
operational way theory and experiment are compared in the
laboratory. Moreover, it is free of paradoxes, which have their
origin in a realist interpretation. As will be seen in the next
section, the difference between realist and empiricist
interpretations is highly relevant when dealing with the EPR
problem.

\section{EPR experiments and Bell experiments}
\label{sec2} 
\begin{figure}[t]
\psfig{figure=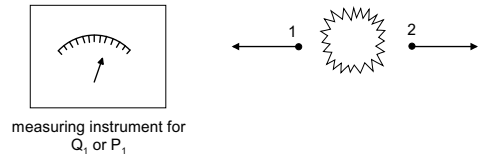,height=1in}
\caption{\protect\label{fig3} EPR experiment.}
\end{figure}

In figure \ref{fig3} the experiment is depicted,
proposed by Einstein, Podolsky and Rosen\cite{EPR} to study
(in)completeness of quantum mechanics. A pair of particles ($1$
and $2$) is prepared in an entangled state and allowed to
separate. A measurement is performed on particle $1$. It is
essential to the EPR reasoning that particle $2$ does {\em not}
interact with any measuring instrument, thus allowing to consider
so-called `elements of physical reality' of this particle, that
can be considered as {\em objective} properties, being
attributable to particle $2$ independently of what happens to
particle $1$. By EPR this arrangement was presented as a way to
perform a measurement on particle $2$ {\em without in any way
disturbing this particle}.

\begin{figure}[t]
\psfig{figure=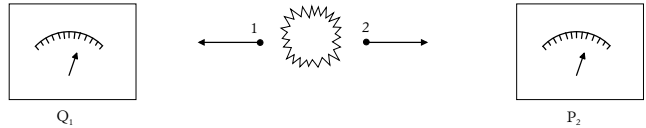,height=.7in}
\caption{\protect\label{fig4} Bell experiment.}
\end{figure}

The EPR experiment should be compared to correlation measurements
of the type performed by Aspect et al.\cite{Asp81,Asp82} to test
Bell's inequality (cf. figure \ref{fig4}).
In these latter experiments also particle $2$ is interacting with
a measuring instrument. In the literature these experiments are
often referred to as EPR experiments, too, thus neglecting the
fundamental difference between the two measurement arrangements of
figures \ref{fig3} and \ref{fig4}. This negligence has been
responsible for quite a bit of confusion, and should preferably be
avoided by referring to the latter experiments as Bell experiments
rather than EPR ones. In EPR experiments particle $2$ is not
subject to a {\em measurement}, but to a (conditional) {\em
preparation} (conditional on the measurement result obtained for
particle $1$). This is especially clear in an empiricist
interpretation, because here measurement results cannot exist
unless a measuring instrument is present, its pointer positions
corresponding to the measurement results.

Unfortunately, the EPR experiment of figure \ref{fig3} was
presented by EPR as a {\em measurement} performed on particle $2$,
and accepted by Bohr as such. That this could happen is a
consequence of the fact that both Einstein and Bohr entertained a
{\em realist} interpretation of quantum mechanical observables
(note that they differed with respect to the interpretation of the
state vector), the only difference being that Einstein's realist
interpretation was an {\em objectivistic} one (in which
observables are considered as properties of the object, possessed
independently of any measurement: the EPR `elements of physical
reality'), whereas Bohr's was a {\em contextualistic} realism (in
which observables are only well-defined within the context of the
measurement). Note that in Bell experiments the EPR reasoning
would break down because, due to the interaction of particle $2$
with its measuring instrument, there cannot exist `elements of
physical reality'.

Much confusion could have been avoided if Bohr had maintained his
{\em interactional} view of measurement. However, by accepting the
EPR experiment as a measurement of particle $2$ he had to weaken
his interpretation to a {\em relational} one (e.g.
Popper\cite{Pop}, Jammer\cite{Jammer}), allowing the observable of
particle $2$ to be co-determined by the measurement context for
particle $1$. This introduced for the first time non-locality in
the interpretation of quantum mechanics. But this could easily
have been avoided if Bohr had required that for a measurement of
particle $2$ a measuring instrument should be actually interacting
with this very particle, with the result that an observable of
particle $n$ ($n=1,2$) can be co-determined {\em in a local way}
by the measurement context of that particle only. This,
incidentally, would have completely made obsolete the EPR
`elements of physical reality', and would have been quite a bit
less confusing than the answer Bohr\cite{Bohr35} actually gave (to
the effect that the definition of the EPR `element of physical
reality' would be ambiguous because of the fact that it did not
take into account the measurement arrangement for the other
particle), thus promoting the non-locality idea.

Summarizing, the idea of EPR non-locality is a consequence of i) a
neglect of the difference between EPR and Bell experiments
(equating `elements of physical reality' to measurement results),
ii) a realist interpretation of quantum mechanics (considering
measurement results as properties of the microscopic object, i.c.
particle $2$). In an {\em empiricist} interpretation there is no
reason to assume any non-locality.

It is often asserted that non-locality is proven by the Aspect
experiments, because these are violating Bell's inequality. The
reason for such an assertion is that it is thought that
non-locality is a necessary condition for a derivation of Bell's
inequality. However, as will be demonstrated in the following,
this cannot be correct since this inequality can be derived from
quite different assumptions. Also, experiments like the Aspect
ones, -although violating Bell's inequality,- do not exhibit any
trace of non-locality, because their measurement results are
completely consistent with the postulate of local commutativity,
implying that relative frequencies of measurement results are
independent of which measurements are performed in causally
disconnected regions. Admittedly, this does not logically exclude
a certain non-locality at the individual level, being unobservable
at the statistical level of quantum mechanical probability
distributions. However, from a physical point of view a peaceful
coexistence between locality at the (physically relevant)
statistical level and non-locality at the individual level is
extremely implausible. Unobservability of the latter would require
a kind of conspiracy not unlike the one making unobservable
19$^{th}$ century world aether. For this reason the `non-locality'
explanation of the experimental violation of Bell's inequality
does not seem to be very plausible, and does it seem wise to look
for alternative explanations.

Since non-locality is never the {\em only} assumption in deriving
Bell's inequality, such alternative explanations do exist. Thus,
Einstein's assumption of the existence of `elements of physical
reality' is such an additional assumption. More generally, in
Bell's derivation\cite{Bell64} the existence of hidden-variables
is one. Is it still possible to derive Bell's inequality if these
assumptions are abolished? Moreover, even assuming the possibility
of hidden-variables theories, are there in Bell's derivation no
hidden assumptions, additional to the locality assumption.

Bell's inequality refers to a set of four quantum mechanical
observables, $A_1,B_1,A_2$ and $B_2$, observables with
different/identical indices being compatible/incompatible. In the
Aspect experiments measurements of the four possible compatible
pairs are performed; in these experiments $A_n$ and $B_n$ refer to
polarization observables of photon $n,\;n=1,2$, respectively).
Bell's inequality can typically be derived for the stochastic
quantities of a classical Kolmogorovian probability theory. Hence,
violation of Bell's inequality is an indication that observables
$A_1,B_1,A_2$ and $B_2$ are not stochastic quantities in the sense
of Kolmogorov's probability theory. In particular, there cannot
exist a quadrivariate joint probability distribution of these four
observables. Such a non-existence is a consequence of the {\em
in}compatibility of certain of the observables. Since
incompatibility is a {\em local} affair, this is another reason to
doubt the `non-locality' explanation of the violation of Bell's
inequality.

In the following derivations of Bell's inequality will be
scrutinized to see whether the non-locality assumption is as
crucial as was assumed by Bell. In doing so it is necessary to
distinguish derivations in quantum mechanics from derivations in
hidden-variables theories.

\section{Bell's inequality in quantum mechanics}
\label{sec3} For dichotomic observables, having values $\pm 1$,
Bell's inequality is given according to
\begin{equation}\label{1}
  |\langle A_1 {A}_2 \rangle - \langle {A}_1 {B}_2
\rangle |-\langle {B}_1 {B}_2 \rangle -\langle {B}_1 {A}_2 \rangle
\leq 2.
\end{equation}
A more general inequality, being valid for arbitrary values of the
observables, is the BCHS inequality
\begin{equation}\label{2}
  -1 \leq p(b_{1}, a_{2}) + p(b_{1}, b_{2}) + p(a_{1}, b_{2}) -
p(a_{1}, a_{2}) - p(b_{1}) - p(b_{2}) \leq 0
\end{equation}
from which (\ref{1}) can be derived for the dichotomic case.
Because of its independence of the values of the observables
inequality (\ref{2}) is preferable by far over inequality
(\ref{1}). Bell's inequality may be violated if some of the
observables are {\em in}compatible:  $[{A}_1,{B}_1]_-\neq
{O},\;[{A}_2,{B}_2]_-\neq {O}$.

I shall now discuss two derivations of Bell's inequality, which
can be formulated within the quantum mechanical formalism, and
which do not rely on the existence of hidden variables. The first
one is relying on a `possessed values' principle, stating that

\noindent
$\left\{\begin{array}{l}\mbox{\em `possessed}\\\mbox{\em values'}\\
\mbox{\em principle}\end{array}\right. =
\left\{\begin{array}{l}\mbox{\em values of quantum mechanical
observables}\\ \mbox{\em may be attributed to the object as }\\
\mbox{\em objective properties, possessed by the object }\\
\mbox{\em independent of observation}\end{array}\right.$

The `possessed values' principle can be seen as an expression of
the objectiv- \nolinebreak istic-realist interpretation of the
quantum mechanical formalism preferred by Einstein (compare the
EPR `elements of physical reality'). The important point is that
by this principle well-defined values are simultaneously
attributed to {\em in}compatible observables. If
$a_i^{(n)},\;b_j^{(n)}=\pm 1$ are the values of $A_i$ and $B_j$
for the $n^{th}$ of a sequence of $N$ particle pairs, then we have
\[
-2\leq a_1^{(n)} a_2^{(n)} -a_1^{(n)}b_2^{(n)}- b_1^{(n)}
b_2^{(n)} - b_1^{(n)} a_2^{(n)} \leq 2,\] from which it directly
follows that the quantities
\[\langle {A}_1 {A}_2 \rangle = \frac{1}{N} \sum^{N}_{n=1} a_1^{(n)}
a_2^{(n)}, \mbox{ \rm etc. } \] must satisfy Bell's inequality
(\ref{1}) (a similar derivation has first been given by
Stapp\cite{Stapp71}, although starting from quite a different
interpretation). The essential point in the derivation is the
assumption of the existence of a quadruple of values
$(a_1,b_1,a_2,b_2)$ for each of the particle pairs.

From the experimental violation of Bell's inequality it follows
that an objectivistic-realist interpretation of the quantum
mechanical formalism, encompassing the `possessed values'
principle, is impossible. Violation of Bell's inequality entails
failure of the `possessed values' principle (no quadruples
available). In view of the important role measurement is playing
in the interpretation of quantum mechanics this is hardly
surprising. As is well-known, due to the incompatibility of some
of the observables the existence of a quadruple of values can only
be attained on the basis of doubtful counterfactual reasoning. If
a realist interpretation is feasible at all, it seems to have to
be a contextualistic one, in which the values of observables are
co-determined by the measurement arrangement. In the case of Bell
experiments non-locality does not seem to be involved.

As a second possibility to derive Bell's inequality within quantum
mechanics we should consider derivations of the BCHS inequality
(\ref{2}) from the existence of a quadrivariate probability
distribution $p(a_{1}, b_{1}, a_{2}, b_{2})$ by Fine\cite{Fine}
and Rastall\cite{Ras83} (also de Muynck\cite{dM86}). Hence, from
violation of Bell's inequality the non-existence of a
quadrivariate joint probability distribution follows. In view of
the fact that incompatible observables are involved, this, once
again, is hardly surprising.

A priori there are two possible reasons for the non-existence of
the quadrivariate joint probability distribution $p(a_{1}, b_{1},
a_{2}, b_{2})$. First, it is possible that $ {\mbox \rm
lim}_{N\rightarrow \infty} N(a_{1}, b_{1}, a_{2}, b_{2})/N$ of the
relative frequencies of quadruples of measurement results does not
exist. Since, however, Bell's inequality already follows from the
existence of relative frequency $N(a_{1}, b_{1}, a_{2}, b_{2})/N$
with {\em finite} $N$, and the limit $N\rightarrow \infty$ is
never involved in any experimental implementation, this answer
does not seem to be sufficient. Therefore the reason for the
non-existence of the quadrivariate joint probability distribution
$p(a_{1}, b_{1}, a_{2}, b_{2})$ can only be the non-existence of
relative frequencies $ N(a_{1}, b_{1}, a_{2}, b_{2})/N$. This
seems to reduce the present case to the previous one: Bell's
inequality can be violated because quadruples
$({A}_1=a_1,{B}_1=b_1,{A}_2=a_2,{B}_2=b_2)$ do not exist.

Could non-locality explain the non-existence of quadruples
$({A}_1=a_1,{B}_1=b_1,{A}_2=a_2,{B}_2=b_2)$? Indeed, it could. If
the value of ${A_1}$, say, is co-determined by the measurement
arrangement of particle $2$, then non-locality could entail
\begin{equation}\label{3}
  a_1({A_2}) \neq a_1({B_2}),
\end{equation}
thus preventing the existence of one single value of observable
${A_1}$ for the two Aspect experiments involving this observable.
This, precisely, is the `non-locality' explanation referred to
above. This explanation is close to Bohr's `ambiguity' answer to
EPR, referred to in section~\ref{sec2}, stating that the
definition of an `element of physical reality' of observable
${A_1}$ must depend on the measurement context of particle $2$.

\begin{figure}[t]
\psfig{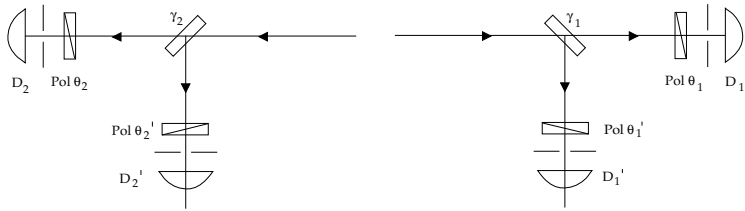}
\caption{\protect\label{fig6} Generalized Aspect experiment.}
\end{figure}

As will be demonstrated next, there is a more plausible {\em
local} explanation, however, based on the inequality
\begin{equation}\label{4}
  a_1({A_1}) \neq a_1({B_1}),
\end{equation}
expressing that the value of ${A_1}$, say, will depend on whether
either ${A_1}$ or ${B_1}$ is measured. Inequality (\ref{4}) could
be seen as an implementation of Heisenberg's disturbance theory of
measurement, to the effect that observables, incompatible with the
actually measured one, are disturbed by the measurement. That such
an effect is really occurring in the Aspect experiments, can be
seen from the generalized Aspect experiment depicted in
figure~\ref{fig6}. This experiment should be compared with the
Aspect switching experiment\cite{Asp82}, in which the switches
have been replaced by two semi-transparent mirrors
(transmissivities $\gamma_1$ and $ \gamma_2$, respectively).
The four Aspect experiments are special cases of the generalized
one, having $\gamma_n=0$ or $1,\; n=1,2$.

Restricting for a moment to one side of the interferometer, it is
possible to calculate the joint detection probabilities of the two
detectors according to
\begin{equation}\label{5}
  (p_{\gamma_1}(a_{1i}, b_{1j})) =
\left( \begin{array}{cc}
0 & \gamma_1 \langle{E^{(1)}}_+\rangle\\
(1-\gamma_1)\langle{F^{(1)}}_+\rangle & 1-\gamma_1
\langle{E^{(1)}}_+\rangle - (1 - \gamma_1)
\langle{F^{(1)}}_+\rangle
\end{array} \right),
\end{equation}
in which $\{{E^{(1)}}_+,{E^{(1)}}_-\}$ and
$\{{F^{(1)}}_+,{F^{(1)}}_-\}$ are the spectral representations of
the two polarization observables (${A_1}$ and ${B_1}$) in
directions $\theta_1$ and $\theta'_1$, respectively. The values
$a_{1i}=+/-,b_{1j}=+/-$ correspond to yes/no registration of a
photon by the detector. $p_{\gamma_1}(+,+)=0$ means that, like in
the switching experiment, only one of the detectors can register
photon $1$. There, however, is a fundamental difference with the
switching experiment, because in this latter experiment the photon
wave packet is sent either toward one detector or the other,
whereas in the present one it is split so as to interact
coherently with both detectors. This makes it possible to
interpret the right hand part of the generalized experiment of
figure~\ref{fig6} as a joint non-ideal measurement of the
incompatible polarization observables in directions $\theta_1$ and
$\theta'_1$ (e.g. de Muynck et al.\cite{dMDBMa95}), the joint
probability distribution of the observables being given by
(\ref{5}).

It is not possible to extensively discuss here the relevance of
experiments of the generalized type for understanding Heisenberg's
disturbance theory of measurement, and its relation to the
Heisenberg uncertainty relations (see e.g. de
Muynck\cite{dM2000}). The important point is that such experiments
do not fit into the standard (Dirac-von Neumann) formalism in
which a probability is an expectation value of a projection
operator. Indeed, from (\ref{5}) it follows that
$p_{\gamma_1}(a_{1i}, b_{1j})=Tr \rho {R^{(1)}}_{ij}$ is yielding
operators ${R^{(1)}}_{ij}$ according to
\begin{equation}\label{6}
({R^{(1)}}_{ ij}) = \left( \begin{array}{cc}
O & \gamma_1 {E^{(1)}}_+\\
(1-\gamma_1){F^{(1)}}_+ & \gamma_1 {E^{(1)}}_- + (1 - \gamma_1)
{F^{(1)}}_- \end{array} \right).
\end{equation}
The set of operators $\{{R^{(1)}}_{ ij}\}$ constitutes a so-called
positive operator-valued measure (POVM). Only generalized
measurements corresponding to POVMs are able to describe joint
non-ideal measurements of incompatible observables. By calculating
the marginals of probability distribution $p_{\gamma_1}(a_{1i},
b_{1j})$ it is possible to see that for each value of $\gamma_1$
information is obtained on both polarization observables, be it
that information on polarization in direction $\theta_1$ gets more
non-ideal as $\gamma_1$ decreases, while information on
polarization in direction $\theta'_1$ is getting more ideal. This
is in perfect agreement with the idea of mutual disturbance in a
joint measurement of incompatible observables. The explanation of
the non-existence of a single measurement result for observable
${A_1}$, say, as implied by inequality (\ref{4}), is corroborated
by this analysis.

The analysis can easily be extended to the joint detection
probabilities of the whole experiment of figure~\ref{fig6}. The
joint detection probability distribution of all four detectors is
given by the expectation value of a quadrivariate POVM
$\{R_{ijk\ell}\}$ according to
\begin{equation}\label{7}
  p_{\gamma_1\gamma_2}(a_{1i}, b_{1j}, a_{2k}, b_{2\ell}) = Tr \rho
{R}_{ijk\ell}.
\end{equation}
This POVM can be expressed in terms of the POVMs of the left and
right interferometer arms according to
\begin{equation}\label{8}
  {R}_{ijk\ell} = {R}^{(1)}_{ ij} {R}^{(2)}_{ k\ell}.
\end{equation}

It is important to note that the existence of the quadrivariate
joint probability distribution (\ref{7}), and the consequent
satisfaction of Bell's inequality, is a consequence of the
existence of quadruples of measurement results, available because
it is possible to determine for each individual particle pair what
is the result of each of the four detectors. Although, because of
(\ref{8}), also locality is assumed, this does not play an
essential role. Under the condition that a quadruple of
measurement results exists for each individual photon pair Bell's
inequality would be satisfied also if, due to non-local
interaction, ${R}_{ijk\ell}$ were not a product of operators of
the two arms of the interferometer.
The reason why the standard Aspect experiments do not satisfy
Bell's inequality is the non-existence of a quadrivariate joint
probability distribution yielding the bivariate probabilities of
these experiments as marginals. Such a non-existence is strongly
suggested by Heisenberg's idea of mutual disturbance in a joint
measurement of incompatible observables. This is corroborated by
the easily verifiable fact that the quadrivariate joint
probability distributions of the standard Aspect experiments,
obtained from (\ref{7}) and (\ref{8}) by taking $\gamma_n$ to be
either $1$ or $0$, are all distinct. Moreover, in general the
quadrivariate joint probability distribution (\ref{7}) for one
standard Aspect experiment does not yield the bivariate ones of
the other experiments as marginals. Although it is not strictly
excluded that a quadrivariate joint probability distribution might
exist having the bivariate probabilities of the standard Aspect
experiments as marginals (hence, different from the ones referred
to above), does the mathematical formalism of quantum mechanics
not give any reason to surmise its existence. As far as quantum
mechanics is concerned, the standard Aspect experiments need not
satisfy Bell's inequality.

\section{Bell's inequality in stochastic and deterministic hidden-variables
 theories}\label{sec4}
In stochastic hidden-variables theories quantum mechanical
probabilities are usually given as
\begin{equation}\label{9}
 p(a_{1}) = \int_\Lambda d\lambda \; \rho(\lambda)
p(a_{1} | \lambda),
\end{equation}
in which $\Lambda$ is the space of hidden variable $\lambda$ (to
be compared with classical phase space), and $p(a_{1} | \lambda)$
is the conditional probability of measurement result $A=a_1$ if
the value of the hidden variable was $\lambda$, and
$\rho(\lambda)$ the probability of $\lambda$. It should be noticed
that expression (\ref{9}) fits perfectly into an empiricist
interpretation of the quantum mechanical formalism, in which
measurement result $a_1$ is referring to a pointer position of a
measuring instrument, the object being described by the hidden
variable. Since $p(a_{1} | \lambda)$ may depend on the specific
way the measurement is carried out, the stochastic
hidden-variables model corresponds to a contextualistic
interpretation of quantum mechanical observables. Deterministic
hidden-variables theories are just special cases in which $p(a_{1}
| \lambda)$ is either $1$ or $0$. In the deterministic case it is
possible to associate in a unique way (although possibly dependent
on the measurement procedure) the value $a_1$ to the phase space
point $\lambda$ the object is prepared in. A disadvantage of a
deterministic theory is that the physical interaction of object
and measuring instrument is left out of consideration, thus
suggesting measurement result $a_1$ to be a (possibly contextually
determined) property of the object. In order to have maximal
generality it is preferable to deal with the stochastic case.

For Bell experiments we have
\begin{equation}\label{10}
p(a_{1}, a_{2}) = \int_\Lambda d\lambda \; \rho(\lambda) p(a_{1},
a_{2} | \lambda),
\end{equation}
a condition of conditional statistical independence,
\begin{equation}\label{11}
p(a_{1}, a_{2} | \lambda) = p(a_{1}| \lambda) p(a_{2} | \lambda),
\end{equation}
expressing that the measurement procedures of $A_1$ and $A_2$ do
not influence each other (so-called locality condition).

As is well-known the locality condition was thought by Bell to be
the crucial condition allowing a derivation of his inequality.
This does not seem to be correct, however. As a matter of fact,
Bell's inequality can be derived if a quadrivariate joint
probability distribution exists\cite{Fine,Ras83}. In a stochastic
hidden-variables theory such a distribution could be represented
by
\begin{equation}\label{12}
p(a_{1},b_{1}, a_{2}, b_{2}) = \int_\Lambda d\lambda \;
\rho(\lambda) p(a_{1},b_{1}, a_{2}, b_{2} | \lambda),
\end{equation}
without any necessity that the conditional probability be
factorizable in order that Bell's inequality be satisfied
(although for the generalized experiment discussed in
section~\ref{sec3} it would be reasonable to require that
$p(a_{1},b_{1}, a_{2}, b_{2} | \lambda)=p(a_{1},b_{1}|
\lambda)p(a_{2}, b_{2} | \lambda)$). Analogous to the quantum
mechanical case, it is sufficient that for each individual
preparation (here parameterized by $\lambda$) a quadruple of
measurement results exists. If Heisenberg measurement disturbance
is a physically realistic effect in the experiments at issue, it
should be described by the hidden-variables theory as well.
Therefore the explanation of the non-existence of such quadruples
is the same as in quantum mechanics.

However, with respect to the possibility of deriving Bell's
inequality there is an important difference between quantum
mechanics and the stochastic hidden-variables theories of the kind
discussed here. Whereas quantum mechanics does not yield any
indication as regards the existence of a quadrivariate joint
probability distribution returning the bivariate probabilities of
the Aspect experiments as marginals, local stochastic
hidden-variables theory does. Indeed, using the single-observable
conditional probabilities assumed to exist in the local theory
(compare (\ref{11})), it is possible to construct a quadrivariate
joint probability distribution according to
\begin{equation}\label{13}
p(a_{1}, a_{2},b_{1}, b_{2}) = \int_\Lambda d\lambda \;
\rho(\lambda) p(a_{1} | \lambda)p(a_{2} | \lambda)p(b_{1} |
\lambda)p(b_{2} | \lambda),
\end{equation}
satisfying all requirements. It should be noted that (\ref{13})
does not describe the results of any joint measurement of the four
observables that are involved. Quadruples $(a_{1}, a_{2},b_{1},
b_{2})$ are obtained here by combining measurement results found
in different experiments, assuming the same value of $\lambda$ in
all experiments. For this reason the physical meaning of this
probability distribution is not clear. However, this does not seem
to be important. The existence of (\ref{13}) as a purely
mathematical constraint is sufficient to warrant that any
stochastic hidden-variables theory in which (\ref{10}) and
(\ref{11}) are satisfied, must require that the standard Aspect
experiments obey Bell's inequality.  Admittedly, there is a
possibility that (\ref{13}) might not be a valid mathematical
entity because it is based on multiplication of the probability
distributions $p(a | \lambda)$, which might be distributions in
the sense of Schwartz' distribution theory. However, the remark
made with respect to the existence of probability distributions as
infinite$-N$ limits of relative frequencies is valid also here:
the reasoning does not depend on this limit, but is equally
applicable to relative frequencies in finite sequences.

The question is whether this reasoning is sufficient to conclude
that no {\em local} hidden-variables theory can reproduce quantum
mechanics. Such a conclusion would only be justified if locality
would be the {\em only} assumption in deriving Bell's inequality.
If there would be any additional assumption in this derivation,
then violation of Bell's inequality could possibly be blamed on
the invalidity of this additional assumption rather than locality.
Evidently, one such additional assumption is the existence of
hidden variables. A belief in the completeness of the quantum
mechanical formalism would, indeed, be a sufficient reason to
reject this assumption, thus increasing pressure on the locality
assumption. Since, however, an empiricist interpretation is hardly
reconcilable with such a completeness belief, we have to take
hidden-variables theories seriously, and look for the possibility
of additional assumptions within such theories.

In expression (\ref{9}) one such assumption is evident, viz. the
existence of the conditional probability $p(a_{1} | \lambda)$. The
assumption of the applicability of this quantity in a quantum
mechanical measurement is far less innocuous than appears at first
sight. If quantum mechanical measurements really can be modeled by
equality (\ref{9}), this implies that a quantum mechanical
measurement result is determined, either in a stochastic or in a
deterministic sense, by an {\em instantaneous} value $\lambda$ of
the hidden variable, prepared independently of the measurement to
be performed later. It is questionable whether this is a realistic
assumption, in particular, if hidden variables would have the
character of rapidly fluctuating stochastic variables. As a matter
of fact, every individual quantum mechanical measurement takes a
certain amount of time, and it will in general be virtually
impossible to determine the precise instant to be taken as the
initial time of the measurement, as well as the precise value of
the stochastic variable at that moment. Hence, hidden-variables
theories of the kind considered here may be too specific.

Because of the assumption of a non-contextual preparation of the
hidden variable, such theories were called {\em
quasi-objectivistic} stochastic hidden-variables theories in de
Muynck and van Stekelenborg\cite{dMvS88} (dependence of the
conditional probabilities $p(a_{1} | \lambda)$ on the measurement
procedure preventing complete objectivity of the theory). In the
past attention has mainly been restricted to quasi-objectivistic
hidden-variables theories. It is questionable, however, whether
the assumption of quasi-objectivity is a possible one for
hidden-variables theories purporting to reproduce quantum
mechanical measurement results. The existence of quadrivariate
probability distribution (\ref{13}) only excludes
quasi-objectivistic local hidden-variables theories (either
stochastic or deterministic) from the possibility of reproducing
quantum mechanics. As will be seen in the next section, it is far
more reasonable to blame quasi-objectivity than locality for this,
thus leaving the possibility of local hidden-variables theories
that are not quasi-objectivistic.

\section{Analogy between thermodynamics and quantum
mechanics} \label{sec5} The essential feature of expression
(\ref{9}) is the possibility to attribute, either in a stochastic
or in a deterministic way, measurement result $a_{1}$ to an {\em
instantaneous} value of hidden variable $\lambda$. The question is
whether this is a reasonable assumption within the domain of
quantum mechanical measurement. Are the conditional probabilities
$p(a_{1}|\lambda)$ experimentally relevant within this domain? In
order to give a tentative answer to this question, we shall
exploit the analogy between thermodynamics and quantum mechanics,
considered already a long time ago by many authors (e.g. de
Broglie\cite{deBroglie95}, Bohm et al.\cite{Bohm53,BoVi},
Nelson\cite{Nelson67,Nelson85}).

\begin{tabular}{ccc}
Quantum mechanics &$\rightarrow$ &Hidden variables theory \\
$({A}_1, {A}_2, {B}_1, {B}_2)$  &&$\lambda$\\
$\updownarrow $&&$\updownarrow$\\
Thermodynamics  &$\rightarrow$ &Classical statistical mechanics\\
$(p,T,S)$& & $\{q_i,p_i\} $\\
\end{tabular}

\noindent In this analogy thermodynamics and quantum mechanics are
considered as phenomenological theories, to be reduced to more
fundamental ``microscopic'' theories. The reduction of
thermodynamics to classical statistical mechanics is thought to be
analogous to a possible reduction of quantum mechanics to
stochastic hidden-variables theory. Due to certain restrictions
imposed on preparations and measurements within the domains of the
phenomenological theories, their domains of application are
thought to be contained in, but smaller than, the domains of the
``microscopic'' theories.

In order to assess the nature and the importance of such
restrictions let us first look at thermodynamics. As is well-known
(e.g. Hollinger and Zenzen\cite{HolZen85}) thermodynamics is valid
only under a condition of molecular chaos, assuring the existence
of local equilibrium\footnote{In ``equilibrium thermodynamics''
equilibrium is assumed to be even global.} necessary for the
ergodic hypothesis to be satisfied. Thermodynamics only describes
measurements of quantities (like pressure, temperature, and
entropy) being defined for such equilibrium states. From an
operational point of view this implies that measurements within
the domain of thermodynamics do not yield information on the
object system, valid for one particular instant of time, but it is
{\em time-averaged} information, time averaging being replaced,
under the ergodic hypothesis, by ensemble averaging. In the Gibbs
theory this ensemble is represented by the canonical density
function $Z^{-1}e^{-H(\{q_n,p_n\})/kT}$ on phase space. This state
is called a {\em macrostate}, to be distinguished from the {\em
microstate} $\{q_n,p_n\}$, representing the point in phase space
the classical object is in at a certain instant of time.

The restricted validity of thermodynamics is manifest in a
two-fold way: i) through the restriction of all possible density
functions on phase space to the canonical ones; ii) through the
restriction of thermodynamical quantities (observables) to
functionals on the space of thermodynamic states. Physically this
can be interpreted as a restriction of the domain of application
of thermodynamics to those measurement procedures probing only
properties of the macrostates. This implies that such measurements
only yield information that is averaged over times exceeding the
relaxation time needed to reach a state of (local) equilibrium.
Thus, it is important to note that thermodynamic quantities are
quite different from the physical quantities of classical
statistical mechanics, the latter ones being represented by
functions of the microstate $\{q_n,p_n\}$ and, hence, referring to
a particular instant of time\footnote{Note that a ``definition''
of an instantaneous temperature by means of the equality $3/2
nkT=\sum_i {\bf p}_i^2/2 m_i $ does not make sense, as can easily
be seen by applying this ``definition'' to an ideal gas in a
container freely falling in a gravitational field.}. Only if it
were possible to perform measurements faster than the relaxation
time, would it be necessary to consider such non-thermodynamic
quantities. Such measurements, then, are outside the domain of
application of thermodynamics. Thus, if we have a cubic container
containing a volume of gas in a microstate initially concentrated
at its center, and if we could measure at a single instant of time
either the total kinetic energy or the force exerted on the
boundary of the container, then these results would not be equal
to thermodynamic temperature and pressure\footnote{Thermodynamic
pressure is defined for the canonical ensemble by $p= kT
\partial/\partial V \log Z$.}, respectively, because this
microstate is not an equilibrium state. Only after the gas has
reached equilibrium within the volume defined by the container
(equilibrium) thermodynamics becomes applicable.

Within the domain of application of thermodynamics the microstate
of the system may change appreciably without the macrostate being
affected. Indeed, a macrostate is equivalent to an (ergodic)
trajectory $\{q_n(t),p_n(t)\}_{ergodic}$. We might exploit as
follows the difference between micro- and macrostates for
characterizing objectivity of a physical theory. Whereas the
microstate is thought to yield an objective description of the
(microscopic) object, the macrostate just describes certain
phenomena to be attributed to the object system only while being
observed under conditions valid within the domain of application
of the theory. In this sense classical mechanics is an objective
theory, all quantities being instantaneous properties of the
microstate. Thermodynamic quantities, only being attributable to
the macrostate (i.e., to an ergodic trajectory), can not be seen,
however, as properties belonging to the object at a certain
instant of time. Of course, we might attribute the thermodynamic
quantity to the event in space-time represented by the trajectory,
but it should be realized that this event is not determined solely
by the preparation of the microstate, but is determined as well by
the macroscopic arrangement serving to define the macrostate.

\begin{figure}[t]
\psfig{figure=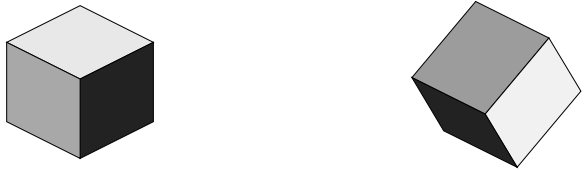,height=.9in}
\caption{\protect\label{fig8} Incompatible thermodynamic
arrangements.}
\end{figure}

In order to illustrate this, consider two identical cubic
containers differing only in their orientations (cf.
figure~\ref{fig8}). In principle, the same microstate may be
prepared in the two containers. Because of the different
orientations, however, the macrostates, evolving from this
microstate during the time the gas is reaching equilibrium with
the container, are different (for different orientations of the
container we have $H_1\neq H_2$, and, hence, $ e^{-H_1/kT}/Z_1
\neq e^{-H_2/kT}/Z_2,$ since $H=T+V$, and $V_1\neq V_2$ because
potential energy is infinite outside a container). This implies
that thermodynamic macrostates may be different even though
starting from the same microstate. Macrostates in thermodynamics
have a {\em contextual} meaning. It is important to note that,
since the container is part of the preparing apparatus, this
contextuality is connected here to {\em preparation} rather than
to measurement. Consequently, whereas classical quantities
$f(\{q_n,p_n\})$ can be interpreted as objective properties,
thermodynamic quantities are non-objective, the non-objectivity
being of a contextual nature.

Let us now suppose that quantum mechanics is related to
hidden-variables theory analogous to the way thermodynamics is
related to classical mechanics, the analogy maybe being even
closer for non-equilibrium thermodynamics (only {\em local}
equilibrium being assumed) than for the thermodynamics of global
equilibrium processes. Support for this idea was found in de
Muynck and van Stekelenborg\cite{dMvS88}, where it was
demonstrated that in the Husimi representation of quantum
mechanics by means of non-negative probability distribution
functions on phase space an analogous restriction to a
``canonical'' set of distributions obtains as in thermodynamics.
In particular, it was demonstrated that the dispersionfree states
$\rho(q,p)= \delta(q-q_0)\delta(p-p_0)$ are not ``canonical'' in
this sense. This implies that within the domain of quantum
mechanics it does not make sense to consider the preparation of
the object in a ``microstate'' with a well-defined value of the
hidden variables $(q,p)$.

In the analogy quantum mechanical observables like ${A}_1, {A}_2,
{B}_1, {B}_2$ should be compared to thermodynamic quantities like
pressure, temperature, and entropy. The central issue in the
analogy is the fact that thermodynamic quantities like pressure
and temperature {\em cannot} be conditioned on the instantaneous
phase space variable $\{q_n, p_n\}$ (microstate). Expressions like
$p(\{q_n, p_n\})$ and $T(\{q_n, p_n\})$ are meaningless within
thermodynamics. Thermodynamic quantities are conditioned on
macrostates, corresponding to ergodic paths in phase space.
Analogously, a quantum mechanical observable might not correspond
to an instantaneous property of the object, but might have to be
associated with an (ergodic) path in hidden-variables space
$\Lambda$ (macrostate) rather than with an instantaneous value
$\lambda$ (microstate).

On the basis of the analogy between thermodynamics and quantum
mechanics it is possible to state the following conjectures:
\begin{itemize}
\item
Quantum mechanical measurements (analogous to thermodynamic
measurements)
 do not probe microstates but macrostates.
\item
Quantum mechanical quantities (analogous to thermodynamic
quantities) should be conditioned on macrostates.
\end{itemize}

A hidden-variables macrostate will be symbolically indicated by
$\overline{\lambda}^t$. For quantum mechanical measurements the
conditional probabilities $p(a_1|\lambda)$ of (\ref{9}) should
then be replaced by $p(a_1|\overline{\lambda}^t)$. Concomitantly,
quantum mechanical probabilities should be represented in the
hidden-variables theory by a functional integral,
\begin{equation}\label{14}
p(a_{1}) = \int d\overline{\lambda}^t \;
\rho(\overline{\lambda}^t) p(a_{1} | \overline{\lambda}^t),
\end{equation}
in which the integration is over all possible macrostates
consistent with the preparation procedure.

By itself conditioning of quantum mechanical observables on
macrostates rather than microstates is not sufficient to prevent
derivation of Bell's inequality. As a matter of fact, on the basis
of expression (\ref{14}) a quadrivariate joint probability
distribution can be defined, analogous to (\ref{13}), according to
\begin{equation}\label{15}
p(a_{1}, a_{2},b_{1}, b_{2}) = \int d\overline{\lambda}^t \;
\rho(\overline{\lambda}^t) p(a_{1} | \overline{\lambda}^t)p(a_{2}
| \overline{\lambda}^t)p(b_{1} | \overline{\lambda}^t)p(b_{2} |
\overline{\lambda}^t),
\end{equation}
from which Bell's inequality can be derived just as well. There
is, however, one important aspect that up till now has not
sufficiently been taken into account, viz. contextuality. In the
construction of (\ref{15}) it is assumed that the macrostate
$\overline{\lambda}^t$ is applicable in each of the measurement
arrangements of observables ${A}_1, {A}_2, {B}_1,$ and $ {B}_2$.
Because of the incompatibility of some of these observables this
is an implausible assumption. On the basis of the thermodynamic
analogy it is to be expected that macrostates
$\overline{\lambda}^t$ will depend on the measurement context of a
specific observable. Since $[{A}_1, {B}_1]_- \neq {O}$, we will
have
\begin{equation}\label{16}
\overline{\lambda}^{t{A_1}} \neq \overline{\lambda}^{t{B_1}},
\end{equation}
and analogously for $A_2$ and $B_2$. Then, for the Bell
experiments measuring the pairs $(A_1,A_2)$ and $(A_1,B_2)$,
respectively, we have
\begin{equation}\label{17}
p(a_1,a_2) =\int d\overline{\lambda}^{tA_1A_2} \;
\rho(\overline{\lambda}^{tA_1A_2}) p(a_{1} |
\overline{\lambda}^{tA_1A_2})p(a_{2} |
\overline{\lambda}^{tA_1A_2}),
\end{equation}
\begin{equation}\label{18}
p(a_1,b_2) =\int d\overline{\lambda}^{tA_1B_2} \;
\rho(\overline{\lambda}^{tA_1B_2}) p(a_{1} |
\overline{\lambda}^{tA_1B_2})p(a_{2} |
\overline{\lambda}^{tA_1B_2}).
\end{equation}
Now, the contextuality expressed by inequality (\ref{16}) prevents
the construction of a quadrivariate joint probability distribution
analogous to (\ref{15}). Hence, like in the quantum mechanical
approach, also in the local non-objectivistic hidden-variables
theory a derivation of Bell's inequality is prevented due to the
local contextuality involved in the interaction of the particle
and the measuring instrument it is directly interacting with.

\section{Conclusions}
\label{sec6} Our conclusion is that if quantum mechanical
measurements do probe macro-\linebreak states $
\overline{\lambda}^{t{A}}$ rather than microstates $\lambda$, then
Bell's inequality cannot be derived {\em for quantum mechanical
measurements}. Both in quantum mechanics and in hidden-variables
theories is Bell's inequality a consequence of the assumption that
the theory is yielding an objective description of reality in the
sense that the preparation of the microscopic object, {\em as far
as relevant to the realization of the measurement result}, can be
thought to be independent of the measurement arrangement. The
important point to be noticed is that, although in Bell
experiments the preparation of the particle pair at the source
(i.e. the microstate) can be considered to be independent of the
measurement procedures to be carried out later (and, hence, one
and the same microstate can be assumed in different Bell
experiments), the measurement result is only determined by the
macrostate, which is co-determined by the interaction with the
measuring instruments. It really seems that the Copenhagen maxim
of the impossibility of attributing quantum mechanical measurement
results to the object as objective properties, possessed
independently of the measurement, should be taken very seriously,
and implemented also in hidden-variables theories purporting to
reproduce the quantum mechanical results. The {\em quantum
mechanical} dice is only cast {\em after} the object has been
interacting with the measuring instrument, even though its result
can be deterministically determined by the (sub-quantum
mechanical) microstate.

The thermodynamic analogy suggests which experiments could be done
in order to transcend the boundaries of the domain of application
of quantum mechanics. If it would be possible to perform
experiments that probe the microstate $\lambda$ rather than the
macrostate $ \overline{\lambda}^{tA}$, then we are in the domain
of (quasi-)objectivistic hidden-variables theories. Because of
(\ref{13}) it, then, is to be expected that Bell's inequality
should be satisfied for such experiments. In such experiments
preparation and measurement must be completed well within the
relaxation time of the microstates. Such times have been estimated
by Bohm\cite{Bohm52} ``for the sake of illustration'' as the time
light needs to cover a distance of the order of the size of an
atom ($10^{-18}$ s, say). If this is correct, then all present-day
experimentation is well within the range of quantum mechanics,
thus explaining the seemingly universal applicability of this
latter theory. By hindsight, this would explain why Aspect's
switching experiment\cite{Asp82} is corroborating quantum
mechanics: the applied switching frequency ($50$ MHz), although
sufficient to warrant locality, has been far too low to beat the
local relaxation processes in each of the measuring instruments
separately.

It has often been felt that the most surprising feature of Bell
experiments is the possibility (in certain states) of a strict
correlation between the measurement results of the two measured
observables, without being able to attribute this to a previous
preparation of the object (no `elements of physical reality ').
For many physicists the existence of such strict correlations has
been reason enough to doubt Bohr's Copenhagen solution to renounce
causal explanation of measurement results, and to replace
`determinism' by `complementarity'. It seems that the urge for
causal reasoning has been so strong that even within the
Copenhagen interpretation a certain causality has been accepted,
even a {\em non}-local one, in an EPR experiment (cf.
figure~\ref{fig3}) determining a measurement result for particle
$2$ by the measurement of particle $1$. This, however, should
rather be seen as an internal inconsistency of this
interpretation, caused by a tendency to make the Copenhagen
interpretation as realist as possible. In a consistent application
of the Copenhagen interpretation to Bell experiments such
experiments could be interpreted as measurements of bivariate
correlation observables. The certainty of obtaining a certain
(bivariate) eigenvalue of such an observable would not be more
surprising than the certainty of obtaining a certain eigenvalue of
a univariate one if the state vector is the corresponding
eigenvector.

It is important to note that this latter interpretation of Bell
experiments takes seriously the Copenhagen idea that quantum
mechanics need not {\em explain} the specific measurement result
found in an individual measurement. Indeed, in order to compare
theory and experiment it would be sufficient that quantum
mechanics just {\em describe} the relative frequencies found in
such measurements. In this view quantum mechanics is just a
phenomenological theory, in an analogous way describing (not
explaining) observations as does thermodynamics in its own domain
of application. Explanations should be provided by ``more
fundamental'' theories, describing the mechanisms behind the
observable phenomena. Hence, the Copenhagen `completeness' thesis
should be rejected (although this need not imply a return to
determinism).

This approach has important consequences. One consequence is that
the non-existence, within quantum mechanics, of `elements of
physical reality' does not imply that `elements of physical
reality' do not exist at all. They could be elements of the ``more
fundamental'' theories. In section~\ref{sec5} it was discussed how
an analogy between quantum mechanics and thermodynamics could be
exploited to spell this out. `Elements of physical reality' could
correspond to hidden-variables microstates $\lambda$. The
determinism necessary to explain the strict correlations, referred
to above, would be explained if, {\em within a given measurement
context}, a microstate would define a unique macrostate
$\overline{\lambda}^{tA}$. This demonstrates how it could be
possible that quantum mechanical measurement results cannot be
attributed to the object as properties possessed prior to
measurement, and there, yet, is sufficient determinism to yield a
local explanation of strict correlations of quantum mechanical
measurement results in certain Bell experiments.

Another important aspect of a dissociation of phenomenological and
fundamental aspects of measurement is the possibility of an {\em
empiricist} interpretation of quantum mechanics. As demonstrated
by the generalized Aspect experiment discussed in
section~\ref{sec3}, an empiricist approach needs a generalization
of the mathematical formalism of quantum mechanics, in which an
observable is represented by a POVM rather than by a
projection-valued measure corresponding to a self-adjoint operator
of the standard formalism. Such a generalization has been very
important in assessing the meaning of Bell's inequality. In the
major part of the literature of the past this subject has been
dealt with on the basis of the (restricted) standard formalism.
However, some conclusions drawn from the restricted formalism are
not cogent when viewed in the generalized one (for instance,
because von Neumann's projection postulate is not applicable in
general). For this reason we must be very careful when accepting
conclusions drawn from the standard formalism. This, in
particular, holds true for the issue
of non-locality.


\section*{References}
\begin{thebibliography}{99}
\bibitem{Heis25}
{W. Heisenberg, {\em Zeitschr. f. Phys.} {\bf 33}, 879 (1925).}

\bibitem{Schroed35}
{E. Schr\"odinger, {\em Naturwissenschaften} {\bf 23}, 807, 823,
844 (1935)
  (English translation in {\em Quantum Theory and Measurement}, eds. J.A.
  Wheeler and W.H. Zurek (Princeton Univ. Press, 1983, p. 152)).}

\bibitem{dM95}
{W.M. de Muynck, {\em Synthese} {\bf 102}, 293 (1995).}

\bibitem{EPR}
{A. Einstein, B. Podolsky, and N. Rosen, {\em Phys. Rev.} {\bf
47}, 777
  (1935).}

\bibitem{Asp81}
{A. Aspect, P. Grangier, and G. Roger, {\em Phys. Rev. Lett} {\bf
47}, 460
  (1981).}

\bibitem{Asp82}
{A. Aspect, J. Dalibard, and G. Roger, {\em Phys. Rev. Lett.} {\bf
49}, 1804
  (1982).}

\bibitem{Pop}
{K.R. Popper, {\em Quantum theory and the schism in physics}
(Rowman and
  Littlefield, Totowa, 1982).}

\bibitem{Jammer}
{M. Jammer, {\em The philosophy of quantum mechanics} (Wiley, New
York, 1974.)}

\bibitem{Bohr35}
{N. Bohr, {\em Phys. Rev.} {\bf 48}, 696 (1935).}

\bibitem{Bell64}
{J.S. Bell, {\em Physics} {\bf 1}, 195 (1964).}

\bibitem{Stapp71}
{H.P. Stapp, {\em Phys. Rev. D} {\bf 3}, 1303 (1971); {\em Il
Nuovo Cim.} {\bf
  29B}, 270 (1975).}

\bibitem{Fine}
{A. Fine, {\em Journ. Math. Phys.} {\bf 23}, 1306 (1982); {\em
Phys. Rev.
  Lett.} {\bf 48}, 291 (1982).}

\bibitem{Ras83}
{P. Rastall, {\em Found. of Phys.} {\bf 13}, 555 (1983).}

\bibitem{dM86}
{W.M. de Muynck, {\em Phys. Lett. A} {\bf 114}, 65 (1986).}

\bibitem{dMDBMa95}
{W.M. de Muynck, W. De Baere, and H. Martens, {\em Found. of
Phys.} {\bf 24},
  1589 (1994).}

\bibitem{dM2000}
{W.M. de Muynck, {\em Found. of Phys.} {\bf 30}, 205 (2000).}

\bibitem{dMvS88}
{W.M. de Muynck and J.T. van Stekelenborg, {\em Ann. der Phys., 7.
Folge}, {\bf
  45}, 222 (1988).}

\bibitem{deBroglie95}
{L. de Broglie, {\em La thermodynamique de la particule isol\'ee}
(Gauthier--Villars, 1964); L. de Broglie, {\em Diverses questions
de m\'ecanique et de thermodynamique classiques et relativistes}
(Springer-Verlag, 1995).}

\bibitem{Bohm53}
{D. Bohm, {\em Phys. Rev.} {\bf 89}, 458 (1953).}

\bibitem{BoVi}
{D. Bohm and J.-P. Vigier, {\em Phys. Rev.} {\bf 96}, 208 (1954).}

\bibitem{Nelson67}
{E. Nelson, {\em Dynamical theories of Brownian motion} (Princeton
University
  Press, 1967).}

\bibitem{Nelson85}
{E. Nelson, {\em Quantum fluctuations} (Princeton University
Press, 1985).}

\bibitem{HolZen85}
{H.B. Hollinger and M.J.Zenzen, {\em The Nature of
Irreversibility} (D. Reidel Publishing Company, Dordrecht, 1985,
sect. 4.4).}

\bibitem{Bohm52}
{D. Bohm, {\em Phys. Rev.} {\bf 85}, 166, 180 (1952).}
\end{thebibliography}
\end{document}